\documentclass[%
10pt,
aps,
prd,
superscriptaddress,
floatfix,
showpacs,
twocolumn,
amsmath,
amssymb,
]{revtex4-2}

\usepackage{graphicx}
\usepackage{subfigure}
\usepackage{url,hyperref} 
\usepackage{color}
\usepackage{bm}

\usepackage{natbib}

\newcommand{\hb}[1]{\hat{\bm{#1}}}
\newcommand{\wt}[1]{\widetilde{#1}}

\begin{document}
\title{Signatures of vacuum birefringence in low-power flying focus pulses}
\author{Martin Formanek}
\email{martin.formanek@eli-beams.eu}
\affiliation{ELI Beamlines Facility, The Extreme Light Infrastructure ERIC, 252 41 Doln\'{i} B\v{r}e\v{z}any, Czech Republic}
\author{John P. Palastro}
\affiliation{Laboratory for Laser Energetics, University of Rochester, Rochester, New York 14623, USA}
\author{Dillon Ramsey}
\affiliation{Laboratory for Laser Energetics, University of Rochester, Rochester, New York 14623, USA}
\author{Stefan Weber}
\affiliation{ELI Beamlines Facility, The Extreme Light Infrastructure ERIC, 252 41 Doln\'{i} B\v{r}e\v{z}any, Czech Republic}
\author{Antonino Di Piazza}
\affiliation{Department of Physics and Astronomy, University of Rochester, Rochester, New York 14627, USA}
\affiliation{Laboratory for Laser Energetics, University of Rochester, Rochester, New York 14623, USA}
\affiliation{Max Planck Institute for Nuclear Physics, Saupfercheckweg 1, D-69117 Heidelberg, Germany}

\date{\today}
%
\begin{abstract}
Vacuum birefringence produces a differential phase between orthogonally polarized components of a weak electromagnetic probe in the presence of a strong electromagnetic field. Despite representing a hallmark prediction of quantum electrodynamics, vacuum birefringence remains untested in pure light configurations due to the extremely large electromagnetic fields required for a detectable phase difference. Here, we exploit the programmable focal velocity and extended focal range of a flying focus laser pulse to substantially lower the laser power required for detection of vacuum birefringence. In the proposed scheme, a linearly polarized x-ray probe pulse counter-propagates with respect to a flying focus pulse, whose focus moves at the speed of light in the same direction as the x-ray probe. The peak intensity of the flying focus pulse overlaps the probe over millimeter-scale distances and induces a polarization ellipticity on the order of $10^{-10}$, which lies within the detection sensitivity of existing x-ray polarimeters.
\end{abstract}
\maketitle

\begin{figure*}
	\includegraphics[width=\textwidth]{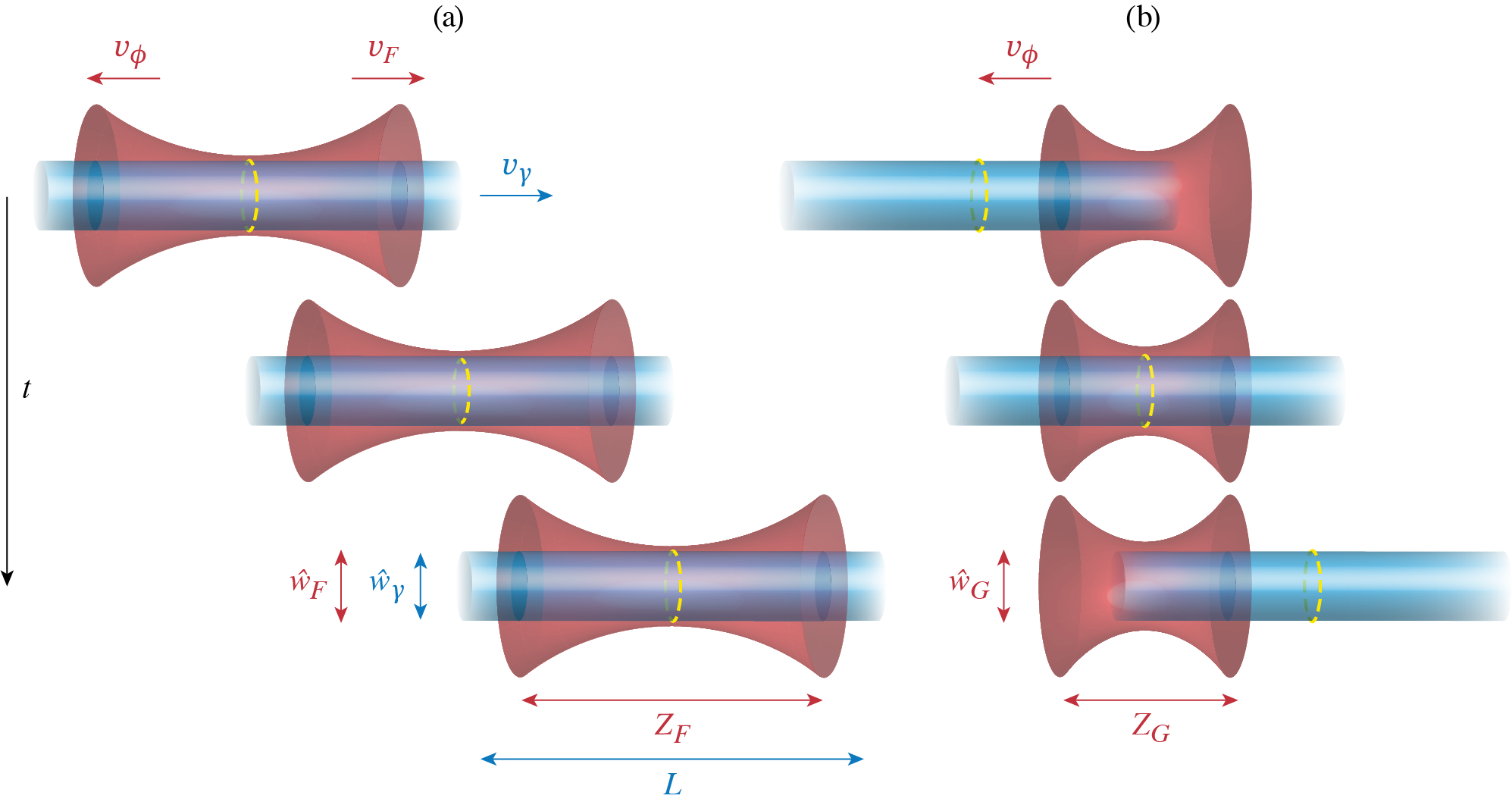}
	\caption{\label{fig:graphics} Interaction geometries for producing an observable signature of vacuum birefringence. An x-ray probe pulse (blue) encounters a (red) co-moving flying focus pulse (a) or conventional Gaussian pulse with a stationary focus (b). The three snapshots from top to bottom show the progression of time. The x-ray probe pulse has a length $L$, spot size $\hat{w}_\gamma$, and travels to the right with a velocity $v_\gamma = c$. The optical pulse has a spot size $\hat{w}_\ell$, Rayleigh range $Z_\ell$, and a phase velocity $v_\phi = -c$, where $\ell$ is either $F$ (FF case) or $G$ (conventional case). The peak-intensity of the FF pulse travels to the right with a velocity $v_F = c$. By extending the interaction length, the FF pulse produces the same birefringent phase difference with a much lower laser power.}
\end{figure*}

\section{Introduction}
Vacuum-polarization effects arise from the interaction of electromagnetic fields in vacuum. These effects are purely quantum in origin, and their discovery contrasts one of the most fundamental principles of classical electrodynamics---the linearity of Maxwell's equations and superposition of their solutions. Well before the foundation of quantum electrodynamics (QED) had been fully developed, it was realized that the existence of antiparticles \cite{Dirac:1928hu} gives rise to nonlinear effects that modify the propagation of electromagnetic waves in vacuum. This idea was first formulated in Refs. \cite{Heisenberg_1936,Weisskopf_1936}, which presented a quantum Lagrangian density for a slowly-varying, but otherwise arbitrary, electromagnetic field that included the effects of electron-positron ``vacuum fluctuations.'' This so-called Euler-Heisenberg (EH) Lagrangian density was later re-derived by Schwinger who employed the proper-time method and techniques of the newly formulated QED \cite{Schwinger_1951}. Within the framework of the EH-Lagrangian, the importance of nonlinear effects depends on the strength of the electromagnetic field as compared to the critical electric and magnetic fields\cite{Heisenberg_1936,Weisskopf_1936,Schwinger_1951}: $E_{cr}=m^2c^3/\hbar|e|\approx 1.3\times 10^{16}\;\text{V/cm}$ and $B_{cr}=m^2c^2/\hbar|e|\approx 4.4\times 10^{9}\;\text{T}$, where $m$ is the electron mass and $e<0$ its charge. 

In order to appreciate the exceedingly large values of the critical fields, they can be compared to some of the most intense electromagnetic fields produced in the laboratory, i.e., those of high-power lasers. The current world record for laser intensity is about $1.1\times 10^{23}\;\text{W/cm$^2$}$ \cite{Yoon_2019}, corresponding to an electric field of about $6.4\times 10^{12}\;\text{V/cm}$. There are several laser facilities, either under construction or planned, that may surpass this record by one-to-two orders of magnitude (e.g., see Refs. \cite{APOLLON_10P,weber2017p3,yoon2022ultra,Bromage_2019,bashinov2014new,gan2021shanghai,peng2021overview} and the Multi-Petawatt Physics Prioritization Workshop report \cite{Di_Piazza_2022}). Nevertheless, the intensities produced at these facilities will still be orders of magnitude below the intensity required to reach the critical fields, $I_{cr}=4.6\times 10^{29}\;\text{W/cm$^2$}$.  

Despite such a large disparity in intensity, vacuum polarization effects can, in principle, be observed at much lower intensities by taking advantage of sensitive detectors, the accumulation of signatures over long interaction lengths, and/or favorable scalings with respect to the frequency of a probe field. These possibilities have led to the proposal of several experimental concepts for detecting vacuum-polarization effects\cite{Euler_1936_a,Akhiezer_1937,Karplus_1951,De_Tollis_1964,Ahmadiniaz_2023,Ahmadiniaz_2023_b}. Examples include harmonic-generation/photon merging and photon splitting in intense laser fields \cite{Di_Piazza_2005,Mendonca:2006az,Narozhny_2007,Brodin_2007,Di_Piazza_2007,Di_Piazza_2013,Gies_2014_b,Sundqvist_2023}, vacuum Bragg scattering and Cherenkov radiation \cite{Kryuchkyan_2011,Macleod_2019,Bulanov_2019,Jirka_2023}, vacuum-polarization effects in plasmas \cite{Di_Piazza_2007_a,Pegoraro_2019,Zhang_2020,Bret_2021,Wan_2022}, and photon-photon scattering in a variety of configurations \cite{Varfolomeev_1966,Lundstroem_2006,Lundin_2007,King_2010,Jeong_2020,Karbstein_2021_c,Dumlu_2022}, among others \cite{Di_Piazza_2012,Dunne_2014,King_2016_c,Karbstein_2020,Gonoskov_2022,Fedotov_2023}. As an alternative to pure light configurations, vacuum-polarization effects have been recently measured in ultra-peripheral heavy-ion collisions \cite{Brandenburg:2022tna}. Of interest here is the vacuum birefringence (and dichroism) experienced by an electromagnetic wave as it propagates through a high-intensity laser beam \cite{Heinzl:2006xc,DiPiazza:2006pr,Tommasini_2010,King_2010_b,Homma_2011,Dinu_2014,Dinu_2014_b,King_2016,Bragin_2017,Gies_2018,Karbstein_2018,King_2018_b,shen2018exploring,Pegoraro_2019_b,Bulanov_2020,Ahmadiniaz_2020,Karbstein_2021,Ataman_2021,Karbstein_2021_b,Ahmadiniaz_2021,Jin_2022,Sainte-Marie_2022,Karbstein_2022,Aleksandrov_2023,Macleod_2023}. 

Previously proposed setups to measure vacuum birefringence employed ultra-intense laser pulses to maximize the observable effects, e.g., the polarization rotation of an x-ray probe, and extremely sensitive diagnostics for their measurement. Here, we propose using the extended focal range and moving focal point of a flying focus (FF) pulse \cite{Sainte-Marie_2017,Froula_2018} to lower the intensity, and power, required to measure vacuum birefringence by orders of magnitude. The first experimental realization of a FF pulse used chromatic focusing of a chirped laser pulse to control the velocity of the focal point over distances much longer than a Rayleigh range \cite{Froula_2018,Turnbull_2018}. More recent concepts would allow for higher focused intensities by using axiparabola-echelon optics \cite{Palastro_2020} or nonlinear media \cite{simpson2020nonlinear,simpson2022spatiotemporal}. These concepts have been motivated by a number of theoretical studies into the unique possibilities that FF pulses offer for laser-based applications and fundamental physics studies \cite{Palastro_2018,Howard_2019,Palastro_2020,Ramsey_2020,Ramsey_2022,Formanek_2023}. In the realm of high-field physics, FF beams have been proposed to enhance observable signatures of the transverse formation length of electromagnetic radiation in the quantum regime \cite{Di_Piazza_2021} and radiation-reaction effects at relatively low laser intensities \cite{Formanek_2022}.

In the context of vacuum-polarization, we consider a setup in which a linearly polarized x-ray probe pulse counterpropagates with respect to a FF pulse. The focus of the FF pulse moves in the same direction and with the same velocity as the probe $v_F = v_\gamma = c$ (left column of Fig. \ref{fig:graphics}). The x-ray probe pulse propagates inside the focus of the FF pulse over a macroscopic distance that is independent of the Rayleigh range $Z_F$ and limited only by the energy of the FF pulse. This is in contrast to configurations that employ conventional fixed-focus Gaussian pulses (right column of Fig. \ref{fig:graphics}), which limit the interaction region to their Rayleigh range $Z_G$. In both cases, vacuum-polarization produces different phase shifts in the two polarization components of the probe pulse, resulting in an ellipticity that accumulates over the interaction length. By extending the interaction length, the FF configuration results in ellipticities that are measurable with state-of-the-art detection techniques at powers (and intensities) that are orders of magnitude lower than those required by conventional laser pulses. 

\section{Phase difference due to vacuum birefringence}
\begin{figure}
	\includegraphics[width=\linewidth]{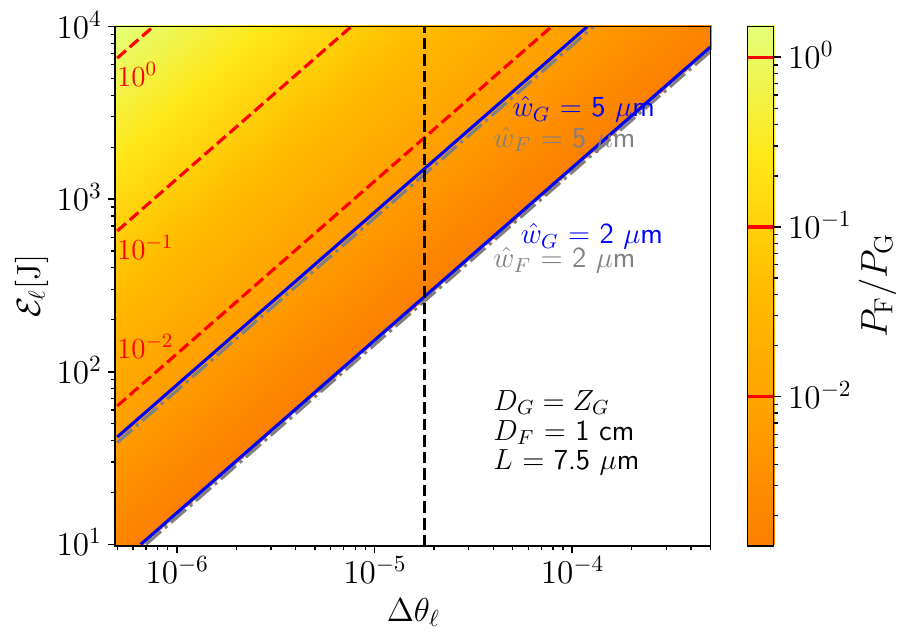}
	\caption{\label{fig:power_ratio} The ratio of laser powers required to achieve a phase difference $\Delta \theta_\ell$ for a given laser pulse energy $\mathcal{E}_\ell$. The dashed red lines correspond to $\log_{10}(P_F/P_G) \in \{-2,-1,0\}$. The solid blue lines and dash-dotted grey lines indicate the spot sizes $\hat{w}_{G/F} = 2$ and 5 $\mu$m for the conventional and FF pulses, respectively. The plot is cut off at $\hat{w}_G = 2\ \mu$m, representing the paraxial approximation limit of a $\lambda_{\ell} = 1\,\mu$m wavelength laser pulse. The vertical dashed line marks the current threshold for experimental detection. For the parameters of the pulses, see the labels and text.}
\end{figure}

The main analytical result of this work is the formula for the phase difference $\Delta \theta_\ell$ between orthogonal polarizations of an x-ray probe pulse induced by a counter-propagating optical laser pulse with energy $\mathcal{E}_\ell$: 
\begin{equation}\label{eq:analytic}
	\Delta \theta_\ell \approx \frac{8\alpha^2}{15\pi}\frac{\mathcal{E}_\ell}{e^2 E_\text{cr}^2} \frac{\hbar \langle \omega_\gamma \rangle}{ \hat{w}_\ell^2}\Sigma_\ell \Lambda_\ell\, ,
\end{equation}
where $\alpha = e^2 / (4\pi\varepsilon_0 \hbar c) \approx 1/137$ is the fine-structure constant and $\langle \omega_\gamma \rangle$ is the average angular frequency of the x-ray pulse. This formula was derived by perturbatively solving Maxwell's equations for an x-ray probe pulse propagating in a medium whose magnetization and polarization correspond to the vacuum modified by an optical laser pulse (see Appendices \ref{sec:app_diff_eq_der}-\ref{sec:app_const_energy} for the detailed derivation). Equation \eqref{eq:analytic} accounts for the transverse structure of the interacting pulses through the factor $\Sigma_\ell$ and the lengths and synchronization of the pulses through the longitudinal form factor $\Lambda_\ell$. Both of these quantities will be discussed below. The spot size of the laser pulse at focus is denoted by $\hat{w}_\ell$, where the subscript $\ell$ is either ``$F$" for a FF pulse or ``$G$" for a conventional Gaussian pulse. Equation \eqref{eq:analytic} allows for a straightforward comparison of FF and conventional Gaussian pulses. A complete description and justification of the approximations that go into the derivation of Eq. \eqref{eq:analytic} is presented in Appendix \ref{sec:app_diff_eq_der}.

Figure \ref{fig:power_ratio} displays the predictions of Eq. \eqref{eq:analytic} and demonstrates that for the same laser pulse energy, a FF pulse can induce the same birefringent phase difference as a conventional Gaussian pulse at a much lower power $P_{\ell}$. The parameters used to generate Fig. \ref{fig:power_ratio} were motivated by current x-ray sources and near-term laser facilities. Specifically, a 10 keV x-ray pulse with a length $L = 7.5\ \mu$m (25 fs) and focal spot $\hat{w}_\gamma = 1.5\ \mu$m colliding with a $\lambda_\ell = 1\ \mu$m wavelength optical pulse. 

The cycle-averaged powers of rectangular FF and conventional pulses are given by 
\begin{equation}\label{eq:powers}
P_{\ell} = \frac{\mathcal{E}_{\ell}}{\tau_{\ell}}, 
\end{equation}
where the pulse duration $\tau_{\ell}$ determines the interaction length: $D_{\ell} = \tau_{\ell}/2$ (see Appendix \ref{sec:app_const_energy}).
In Fig. \ref{fig:power_ratio}, the interaction length of the FF pulse was chosen to be $D_F = 1$ cm based on experimentally demonstrated focal ranges \cite{Froula_2018}. Perfect synchronization between the center of the x-ray probe pulse and intensity peak of the FF pulse was also assumed. The interaction length of the conventional pulse was set to $D_G = Z_G$, where $Z_G \equiv \omega_G \hat{w}_G^2/2c$ is its Rayleigh range and $\omega_G$ its angular frequency. The centers of the conventional and x-ray pulses were set to meet at the focus of the conventional pulse (see Appendix \ref{sec:perturbative}). This ensures a near-optimal configuration where the conventional and x-ray pulses interact over an entire Rayleigh range of the conventional pulse. For fixed laser energy and phase difference, the spot sizes of the conventional and FF pulses are nearly equal (Fig. \ref{fig:power_ratio}), such that $\Delta \theta_{\ell} \propto \mathcal{E}_{\ell} \propto P_FD_F \approx P_GD_G$, or
\begin{equation}\label{eq:powers2}
P_F \approx \frac{D_G}{D_F}P_G. 
\end{equation}
Thus the FF reduces the power required for an observable phase difference by extending the interaction length: $D_F \gg D_G$ implies $P_F \ll P_G$. 


The vertical dashed line in Fig. \ref{fig:power_ratio} indicates the threshold for currently measurable phase differences, i.e., $\Delta \theta_{\ell} = 1.8\times 10^{-5}$. The ellipticity of the x-ray pulse $\delta^2$ is related to the phase difference by $\delta^2 \approx \Delta \theta_{\ell}^2/4$. It is assumed that the x-ray pulse is initially linearly polarized at an angle of 45$^\circ$ with respect to the fields of the optical pulse and has polarization purity better than the detection threshold. This can be achieved using a monochromator and multiple Bragg reflections from channel-cut crystals \cite{schulze2022towards} and tested by a null experiment without the laser pulse. A narrow band x-ray pulse ($\sim$ 1 eV bandwidth \cite{schulze2022towards}) is produced. Ellipticities $\delta^2 \approx 8\times 10^{-11}$ are within the detection limits of existing experimental techniques \cite{Heinzl:2006xc,schulze2022towards}. As an example, a 1 kJ laser pulse focused to a $\hat{w}_\ell = 3\ \mu$m spot can induce a phase difference $\Delta \theta_{\ell} = 3.5 \times 10^{-5}$ and an ellipticity $\delta^2 \approx 3\times10^{-10}$. In the conventional case, this would require a power of 5.3 PW and an intensity of $7.8 \times 10^{22}$ W/cm$^2$, which is  nearly equal to the world record \cite{Yoon_2019}. In the FF case, only 15 TW are needed, corresponding to an intensity of $2.2\times 10^{20}$ W/cm$^2$, a value approximately 350 times smaller than the conventional case.

\begin{figure}
	\includegraphics[width=\linewidth]{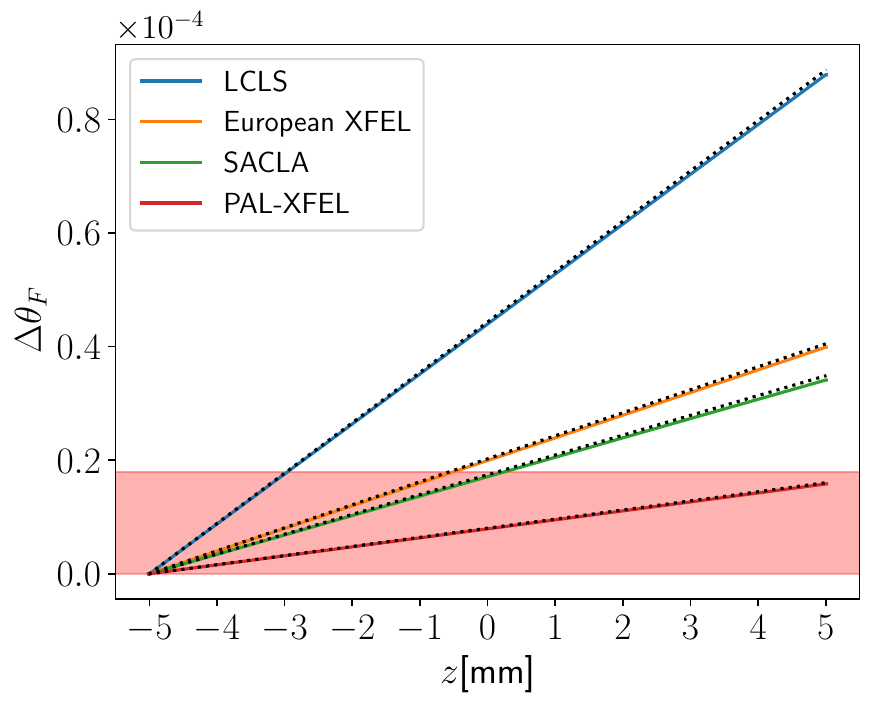}
	\caption{\label{fig:birefringence} Accumulated phase difference between orthogonal polarizations of an x-ray pulse from an XFEL interacting with a FF pulse as a function of distance relative to the focus of the x-ray pulse ($z=0$). The FF pulse has a $\lambda_F = 1\ \mu$m wavelength, $\hat{w}_F = 3\ \mu$m spot size, $P_F = 15$ TW, and $\mathcal{E}_F$ = 1 kJ. See Table \ref{tab:xfels} for the XFEL parameters. Phase differences outside of the red shaded region would be measurable with current x-ray polarimeters \cite{Heinzl:2006xc,schulze2022towards}. The analytical results are plotted as dotted lines.}   
\end{figure}

\begin{table}
	\begin{tabular}{c|ccccc}
		\hline \hline
		Name	&	$\hbar\omega_\gamma$[keV]	&	$\lambda_\gamma$[nm]	&	$\hat{w}_\gamma$[$\mu$m]	& $L$[fs] &	Ref.\\
		\hline
		LCLS    & 25 & 0.0496 & 1 & 50 & \cite{lcls}\\
		European XFEL	&	15	&	0.0827	&	3	&	25& \cite{euxfel}	\\
		SACLA	&	10	&	0.124	&	1.4	&	10 &\cite{matsuyama2016nearly}	\\
		PAL-XFEL	&	9.7	&	0.128	&	5	& 25 &\cite{palxfel}	\\
		\hline \hline 
	\end{tabular}
	\caption{\label{tab:xfels}Parameters of the x-ray probe pulses used in the analytical estimates and numerical simulations for Fig. \ref{fig:birefringence}.}
\end{table}

These examples suggest that vacuum birefringence measurements could be  made experimentally accessible by pairing a high-energy laser system capable of producing a FF pulse with a hard x-ray source. Figure \ref{fig:birefringence} shows the accumulated phase difference $\Delta \theta_F$ that could be achieved with x-ray pulses from currently available XFELs (see Tab. \ref{tab:xfels} for the parameters). For each XFEL considered, the phase difference reaches measurable values after just a few millimeters of propagation (outside the red shaded region).  

Ultimately, the experimental feasibility of measuring the phase difference will depend on the sensitivity of $\Delta \theta_{\ell}$ to the transverse and longitudinal overlap of the optical and x-ray pulses. In Fig. \ref{fig:power_ratio}, near-ideal longitudinal overlap was assumed. For the FF, this means that the longitudinal center of the x-ray pulse was co-located with the peak intensity of the FF. For the conventional pulse, this means that the leading and trailing edges of the x-ray and optical pulses met symmetrically at points located a distance $Z_G/2$ from the focus of the optical pulse.

\begin{figure}
	\subfigure{\includegraphics[width=\linewidth]{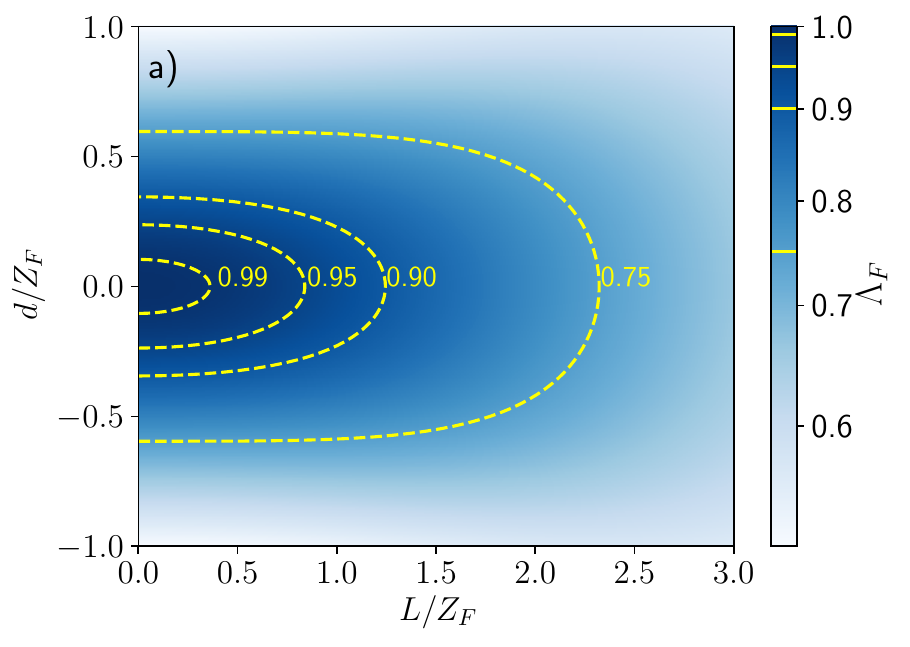}\label{fig:LambdaF}}\\
	\subfigure{\includegraphics[width=\linewidth]{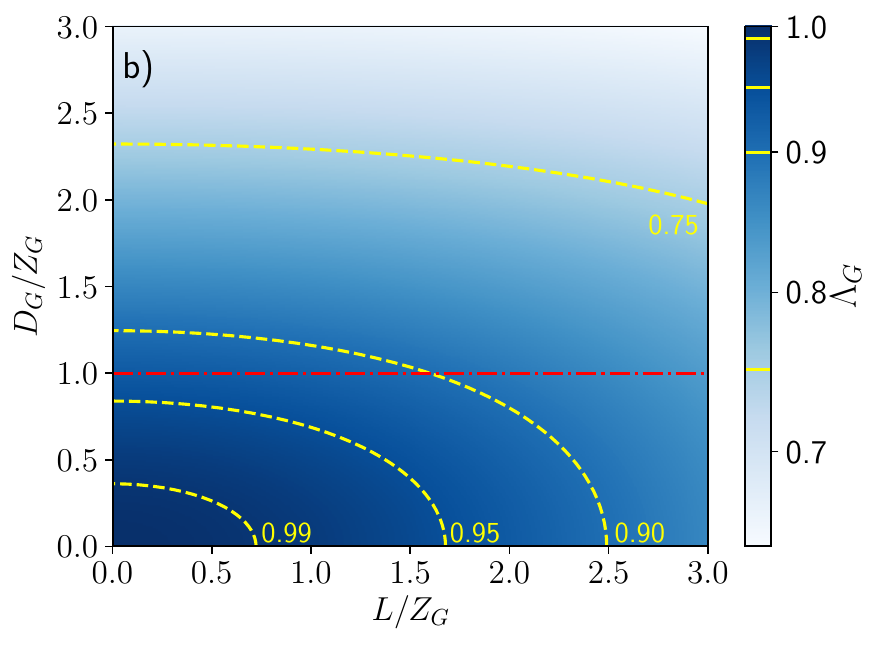}\label{fig:LambdaG}}
	\caption{\label{fig:numerical_factor} Geometric form factor $\Lambda_\ell$ for the FF configuration (a) and a conventional configuration (b). The x-ray probe pulse has a length $L$. In the FF configuration, the x-ray pulse is offset from the center of the FF intensity peak by a distance $d$. In the conventional configuration, the interaction length is $D_G$. For this plot, the spot sizes are $\hat{w}_\gamma = 1.5\ \mu$m and $\hat{w}_{\ell} = 3\ \mu$m. The red dash-dotted line indicates $D_G = Z_G$.}
\end{figure}

The effect of imperfect longitudinal overlap on $\Delta \theta_{\ell}$ is captured by the form factor $\Lambda_{\ell}$ and is illustrated in Fig. \ref{fig:numerical_factor}. The full analytic expressions for the $\Lambda_{\ell}$ appear in Appendices \ref{sec:perturbative} and \ref{sec:app_const_energy}. The expressions can be summarized as follows. In the FF case, $\Lambda_F$ depends on the length of the x-ray probe pulse $L$ and the offset of its center from the traveling intensity peak of the FF $d$ [Fig. \ref{fig:LambdaF}]. For ultrashort x-ray pulses ($L \ll Z_F$) that are colocated with the center of the FF intensity peak ($d = 0$), $\Lambda_{F} \approx 1$. X-ray pulses that are either offset from the center of the FF intensity peak or appreciably longer than the Rayleigh range of the FF will experience a lower intensity. This causes a smaller value of the form factor and, as a result, the phase difference.

In the conventional case, $\Lambda_G$ depends on the length of the x-ray probe pulse $L$ and on the interaction length $D_G$ [Fig. \ref{fig:LambdaG}]. When $D_G \ll Z_G$, an ultrashort x-ray pulse ($L \ll Z_G$) will encounter an approximately constant field amplitude near the center of the fixed focus, such that $\Lambda_G \approx 1$. The near-optimal case of $D_G = Z_G$ used in Fig. \ref{fig:power_ratio} is displayed as the red dash-dotted line in Fig. \ref{fig:LambdaG}. For longer interactions lengths or x-ray pulse lengths, the x-ray pulse intersects the conventional pulse while it is out of focus and has a lower intensity. This decreases the form factor and results in a smaller phase difference.

The effect of transverse overlap is captured by the factor $\Sigma_\ell$, which depends solely on the ratio of the x-ray and optical spot sizes $\sigma_\ell \equiv \hat{w}_\gamma / \hat{w}_\ell$:
\begin{equation}
	\Sigma_\ell \equiv \frac{1 + 2 \sigma^2_\ell}{(1 + \sigma_\ell^2)^2}\,.
\end{equation}
As $\sigma_\ell \rightarrow 0$, $\Sigma_\ell$ rapidly approaches $1$. Thus, it is unnecessary to focus the x-ray pulse to a spot size that is much smaller than that of the laser pulse. For instance, when the spot size of the x-ray pulse is half that of the laser pulse, $\sigma_\ell = 0.5$ and $\Sigma_\ell = 0.96$. Aside from being easier to realize in practice, larger x-ray spot sizes improve the validity of the analytical approximation $D_F/Z_\gamma \ll 1$ (see Appendix \ref{sec:app_diff_eq_der}) used to derive Eq. \eqref{eq:analytic}.

In optimal conditions where $\Sigma_F = \Sigma_G = \Lambda_F = \Lambda_G = 1$, FF and conventional pulses with equal energies result in identical phase differences. This means that FF pulses do not provide any enhancement in the phase difference when compared to experimentally relevant ultrashort Gaussian pulses. However, the FF configuration requires significantly lower laser powers and peak intensities, allowing for more controllable conditions at the cost of proportionally longer interaction lengths.

\section{Summary and Conclusions}

The extreme field scales inherent to nonlinear QED complicate experimental efforts to test hallmark predictions of the theory, such as vacuum polarization and birefringence. In order to produce an observable signature that could test these predictions, an experimental configuration must take advantage of strong or high-frequency fields, long interaction lengths, or sensitive detectors. A promising configuration for detecting vacuum birefringence uses the collision of a conventional, high-intensity laser pulse with an x-ray probe pulse to induce a differential phase between orthogonally polarized components of the x-ray pulse. However, even in this configuration, an extremely high laser intensity is required to produce an appreciable phase difference. This is because the interaction length is limited by the Rayleigh range of the laser pulse. 

The programmable focal velocity and extended focal range of a flying focus (FF) pulse allows for the accumulation of the birefringent phase difference over much longer distances. This reduces the required laser intensity (and power) by orders of magnitude. Unlike a conventional Gaussian pulse, the interaction length of a FF pulse is independent of the Rayleigh range. Thus the interaction length can be increased without changing the spot size. As a result, a FF pulse with the same energy as a conventional pulse can induce the same birefringent phase difference at a much lower power and intensity. 

Specific examples were presented for kJ-class laser pulses colliding with x-ray pulses from current XFEL facilities. In one such example, detectable signatures of vacuum birefringence were possible with either a 5.3 PW conventional pulse or a 15 TW FF pulse. By mitigating the need for temporal compression to achieve a high laser power, the FF configuration could alleviate engineering constraints on the optical elements. Further, lower laser powers (and intensities) allow for more reliable {\it in situ} diagnostics of the pulse. 

The focal range of the FF pulse used in the examples was 1 cm, which has already been experimentally demonstrated \cite{Froula_2018}, albeit  at low intensities ($10^{14}$ W/cm$^2$). Several paths to higher intensities have been outlined in Refs. \cite{Palastro_2020,simpson2020nonlinear,simpson2022spatiotemporal}. Kilojoule class short pulse systems, such as the OMEGA-EP laser \cite{maywar2008omega}, are in routine operation, and tight focusing is regularly used at several facilities \cite{Yoon_2019,tiwari2019beam,gales2018extreme,obst2023high}. 
The promise of this work, and others like it, will continue to motivate the technological development of FF pulses at higher intensities. Once realized, the configuration proposed here would provide a unique experimental platform for testing properties of the QED vacuum. For instance, the detectable signatures of other effects arising from the EH Lagrangian, such as photon splitting and photon-photon scattering \footnote{Please note, that vacuum birefringence is the phase difference between orthogonally polarized components of the scattered photons of the x-ray probe pulse, whereas photon-photon scattering typically refers to the change in the wavevector of the probe photons.}, also accumulate with the interaction distance. For photon-photon scattering in particular, the framework developed in this paper could be adapted to look at the defocusing of the probe pulse instead. These effects can be experimentally distinguished from the effect discussed here by detectors with frequency or angular discrimination, the details of which will be left to future studies.

\begin{acknowledgments}
	M.F. would like to thank Eric Galtier for useful discussion about XFEL sources. This project has received funding from the European Union’s Horizon Europe research and innovation program under the Marie Sk\l{}odowska-Curie grant agreement No. 101105246-STEFF. This publication is also supported by Collaborative Research Centre 1225 funded by Deutsche Forschungsgemeinschaft (DFG, German Research Foundation)–Project No. 273811115–SFB 1225. The work of J.P.P. and D.R. is supported by the Office of Fusion Energy Sciences under Award Number DE-SC0021057, the Department of Energy National Nuclear Security Administration under Award Number DE-NA0003856, the University of Rochester, and the New York State Energy Research and Development Authority.
	
	In the finalization phase of our manuscript, the paper ``Enhancement of vacuum birefringence with pump laser of flying focus'' was published in Phys. Rev. A \cite{Jin:2023}. Although the authors study the same problem that is considered here, the theoretical approaches and focus are different, e.g., we compare the birefringent signals produced by optimal conventional and flying focus pulses with the same energy.
\end{acknowledgments}

\appendix
\section{Wave equation for the x-ray pulse}\label{sec:app_diff_eq_der}
The vacuum response to external electromagnetic fields is described by the Euler-Heisenberg (EH) Lagrangian density \cite{Heisenberg_1936,Weisskopf_1936,Schwinger_1951}. This density was computed in the limit of uniform and constant fields, i.e., it does not depend on their spacetime derivatives. The characteristic scale for the space (time) variations is determined by the reduced Compton wavelength (Compton time) $\lambdabar_C=\hbar/mc\approx 3.9\times 10^{-11}\;\text{cm}$ ($\lambdabar_C/c=\hbar/mc^2\approx1.3\times 10^{-21}\;\text{s}$), which is about six orders of magnitude smaller than the characteristic wavelength and period of optical laser pulses. As a result, the derivative corrections to the EH Lagrangian will be ignored. Until Appendix \ref{sec:app_const_energy} the units $\hbar=\varepsilon_0 = c = 1$ are used for the convenience of derivation. 

To illustrate the validity of Eq. (\ref{eq:analytic}), consider the underlying assumptions of the model: 

\begin{enumerate}
    \item\label{app:parax} The paraxial approximation is valid for both the  x-ray and laser pulse. For the x-ray pulse this is a reasonable assumption because the neglected longitudinal field is suppressed by a factor $\lambda_{\gamma}/ 2\pi\hat{w}_{\gamma}$ compared to the transverse components. Corrections to the transverse field are suppressed by the square of this factor. For 10 keV pulse focused to $\hat{w}_\gamma = 1\ \mu$m this factor is $2\times 10^{-5}$. For the laser pulse the transverse field correction is zero on axis and suppressed by the factor $(r^2\lambda_\ell^2) / 2\pi \hat{w}_\ell^4$ off axis. As a result, this correction is negligible when either a) the focused spot size $\hat{w}_\ell > 2 \lambda_\gamma$ or b) the focused spot size of the x-ray pulse $\hat{w}_{\gamma} < \hat{w}_{\ell}$. The longitudinal field of the laser pulse can be neglected outright, because it only contributes to the birefringence calculation when multiplied by the negligible longitudinal field of the x-ray pulse.
    \item\label{app:mono} Only frequencies composing the initial x-ray pulse are detected. This can be achieved experimentally by using a spectral filter or spectrometer.
    \item\label{app:svea} The period (wavelength) of the x-ray probe pulse is much shorter than its duration (length), i.e.,  the 
    slowly varying envelope approximation (SVEA) can be applied. For a typical 10 fs long 10 keV x-ray pulse, the period is four orders of magnitude smaller than the duration.
    \item\label{app:weak} The nonlinear vacuum response is small, and the solutions need only be found to first order in the field intensity $\propto \hat{E}_\ell^2 / E_\text{cr}^2$. This is justified for laser intensities much lower than the critical intensity $\sim$$10^{29}\;\text{W/cm$^2$}$, which is certainly the case for any available or foreseeable laser pulse.
    \item\label{app:gaussian} The x-ray probe pulse can be modeled as a Gaussian beam with a rectangular temporal profile. This assumption holds because the x-ray probe pulse does not need to be tightly focused. Other temporal profiles can be accommodated by averaging over intensity. 
	\item\label{app:short} The interaction length $D_\ell$ is much smaller than the Rayleigh range of the x-ray probe pulse $Z_\gamma$. This is readily satisfied for conventional, high-intensity Gaussian pulses, which are typically only tens of femtoseconds (tens of microns) long. With FF pulses, on the other hand, the interaction can be sustained over millimeter-scale distances. Thus, depending on the interaction length, higher x-ray photon energies or more weakly focused x-ray pulses may be required. For all FF and x-ray parameters considered here, the assumption is valid.
	\item\label{app:rect2} The temporal profiles of the laser pulses can be approximated as rectangles. It is primarily the total energy of the optical pulse, not its temporal profile, that determines $\Delta \theta_{\ell}$ (see Appendix \ref{sec:app_numerical}).
\end{enumerate}
These assumptions are referenced in the Appendices \ref{sec:app_diff_eq_der}-\ref{sec:perturbative} and when applicable, the order of neglected terms is included. The last two assumptions have only been made to simplify the analytical calculations. Numerical simulations that make neither of these assumptions are in excellent agreement with Eq \eqref{eq:analytic}, see Appendix \ref{sec:app_numerical}. 

The wave equation for the electric field of the x-ray probe pulse in the medium (polarized vacuum in this case) is derived from Maxwell equations and is given by \cite{Jackson_b_1975}
\begin{equation}\label{eq:hemlhotz}
	(\nabla^2 - \partial_t^2)\bm{E}_\gamma = 4\pi[\nabla \times \partial_t \bm{M} + \partial_t^2 \bm{P} - \nabla(\nabla \cdot \bm{P})]\,.
\end{equation}
The polarization $\bm{P}$ and the magnetization $\bm{M}$ of the quantum vacuum are obtained from the EH Lagrangian as \cite{DiPiazza:2006pr,Di_Piazza_2012}
\begin{align}
	\label{eq:P}\bm{P} &= \frac{\alpha}{180\pi^2 E_\text{cr}^2}[2(E^2-B^2)\bm{E} + 7(\bm{E}\cdot\bm{B})\bm{B}]\,,\\
	\label{eq:M}\bm{M} &= \frac{\alpha}{180\pi^2 E_\text{cr}^2}[2(B^2-E^2)\bm{B} + 7(\bm{E}\cdot\bm{B})\bm{E}]\,,
\end{align}
where $\bm{E} = \bm{E}_\ell + \bm{E}_\gamma$ and $\bm{B} = \bm{B}_\ell + \bm{B}_\gamma$  are the combined electric and magnetic fields of the intense laser and probe pulses, respectively \footnote{Please note a typo in the expression of the magnetization in Ref. \cite{DiPiazza:2006pr}.}. In the configuration considered here, the laser field is polarized in the positive $\hb{x}$ direction and has a phase velocity in the negative $\hb{z}$ direction 
\begin{equation}
	\label{eq:ansatzell1}\begin{split}
	\bm{E}_{\ell}(t,\bm{x}) &= \frac{E_\ell(t,\bm{x})}{2} e^{-i\omega_\ell(t+z)}\hb{x}\\
    &+ \text{c.c.} + \mathcal{O}\left(1/\omega_\ell \hat{w}_\ell\right)\,,
    \end{split}
\end{equation}
\begin{equation}
	\label{eq:ansatzell2}\begin{split}
	\bm{B}_{\ell,\perp}(t,\bm{x}) &= -\frac{E_\ell(t,\bm{x})}{2} e^{-i\omega_\ell(t+z)} \hb{y}\\
    &+ \text{c.c.} + \mathcal{O}\left(1/\omega_\ell\hat{w}_\ell\right)\,,
    \end{split}
\end{equation}
where $E_\ell(t,\bm{x})$ is the slowly-varying, complex envelope and $\omega_\ell = 2\pi/\lambda_\ell$ is laser frequency. The x-ray probe pulse is polarized in the $\hb{x}$ and $\hb{y}$ directions and has a phase velocity in positive $\hb{z}$ direction
\begin{equation}
	\label{eq:ansatzgamma1}\begin{split}\bm{E}_\gamma(t,\bm{x}) &= \frac{E_{\gamma,x}(t,\bm{x})\hb{x} + E_{\gamma,y}(t,\bm{x})\hb{y}}{2}e^{-i\omega_\gamma(t-z)} \\
    &+ \text{c.c.}+ \mathcal{O}(1/\omega_\gamma \hat{w}_\gamma)\,,\end{split}
\end{equation}
\begin{equation}
	\label{eq:ansatzgamma2}\begin{split}\bm{B}_\gamma(t,\bm{x}) &= \frac{E_{\gamma,x}(t,\bm{x})\hb{y} - E_{\gamma,y}(t,\bm{x})\hb{x}}{2}e^{-i\omega_\gamma(t-z)}\\
    & + \text{c.c.}+ \mathcal{O}(1/\omega_\gamma \hat{w}_\gamma) \,,\end{split}
\end{equation}
where the $E_{\gamma,j}(t,\bm{x})$ are the slowly-varying, complex envelopes for the orthogonal polarization components and $\omega_\gamma = 2\pi/\lambda_\gamma$ is the x-ray photon frequency. 

Equations \eqref{eq:ansatzell1}-\eqref{eq:ansatzgamma2} employ the paraxial approximation (Approx. \ref{app:parax}) and include only the dominant electric field components, i.e., those in the polarization directions. With these prescriptions, the field invariants $\mathcal{F}=(1/2)(B^2-E^2)$ and $\mathcal{G} = -\bm{E}\cdot \bm{B}$ are identically zero when evaluated using the laser and x-ray fields independently. However, cross-terms in the invariants, which correspond to interactions of the fields, are non-zero because the phase velocities of the two pulses are equal and opposite. These cross terms provide the largest contribution to the vacuum birefringence experienced by the x-ray pulse.

The exact fields of a conventional Gaussian or FF beam have non-zero field invariants off axis (see Supplemental Material of \cite{Formanek_2022}). While this would introduce additional terms in $\bm{P}$ and $\bm{M}$, these terms are either (1) too small to significantly affect the propagation of the laser pulse or (2) negligible compared to the dominant terms that affect the propagation of the x-ray pulse. More specifically, these terms are $\sim \lambda_\ell^2/(4\pi^2e_{E}\hat{w}_\ell^2)$ times smaller than the dominant contributions \cite{Formanek_2022}, where $e_E = 2.718$ is Euler's number. This factor is approximately $10^{-3}$ for the parameters considered here, $\hat{w}_\ell = 3 \lambda_\ell$. Therefore the non-zero contributions to the individual invariants will be neglected, and $\bm{P}$ and $\bm{M}$ will only include the dominant contributions to the invariants from the interaction terms. 

Using Eqs. \eqref{eq:ansatzell1} - \eqref{eq:ansatzgamma2}, the interaction terms in the invariants are given by
\begin{widetext}
\begin{align}
	E^2 - B^2 &= 4 \left(\frac{E_\ell(t,\bm{x})}{2} e^{-i\omega_\ell(t+z)} + \text{c.c.} \right)\left(\frac{E_{\gamma,x}(t,\bm{x})}{2} e^{-i\omega_\gamma(t-z)} + \text{c.c}\right) + \mathcal{O}(\nu)\,,\\
	\bm{E}\cdot\bm{B} &= -2 \left(\frac{E_\ell(t,\bm{x})}{2} e^{-i\omega_\ell(t+z)} + \text{c.c.} \right)\left(\frac{E_{\gamma,y}(t,\bm{x})}{2} e^{-i\omega_\gamma(t-z)} + \text{c.c}\right) + \mathcal{O}(\nu) \,.
\end{align}
where $\nu \equiv \text{max} (1/\omega_\gamma\hat{w}_\gamma, 1/\omega_\ell \hat{w}_\ell)$. Then, to linear order in the probe field
\begin{equation}
    \begin{split}
	\bm{P}_{\omega_\gamma}(t,\bm{x}) &= \frac{1}{2\pi}\left[\eta_x \left(\frac{E_{\gamma,x}(t,\bm{x})}{2} e^{-i\omega_\gamma(t-z)} + \text{c.c}\right)\hb{x}+ \eta_y \left(\frac{E_{\gamma,y}(t,\bm{x})}{2} e^{-i\omega_\gamma(t-z)} + \text{c.c}\right)\hb{y}\right]\\
 &\times \left(\frac{E_\ell(t,\bm{x})}{2} e^{-i\omega_\ell(t+z)} + \text{c.c.} \right)^2 + \mathcal{O}(\nu)\,,
    \end{split}
 \end{equation}
 \begin{equation}
    \begin{split}
    \bm{M}_{\omega_\gamma}(t,\bm{x}) &= \frac{1}{2\pi} \left[\eta_x \left(\frac{E_{\gamma,x}(t,\bm{x})}{2} e^{-i\omega_\gamma(t-z)} + \text{c.c}\right)\hb{y}-\eta_y \left(\frac{E_{\gamma,y}(t,\bm{x})}{2} e^{-i\omega_\gamma(t-z)} + \text{c.c}\right)\hb{x}\right]\\
    &\times \left(\frac{E_\ell(t,\bm{x})}{2} e^{-i\omega_\ell(t+z)} + \text{c.c.} \right)^2 + \mathcal{O}(\nu) \,,
    \end{split}
\end{equation}
\end{widetext}
where 
\begin{equation}
    \eta_x \equiv \frac{4\alpha}{45\pi E_\text{cr}^2}\,, \quad \eta_y \equiv \frac{7\alpha}{45 \pi E_\text{cr}^2}\,.
\end{equation}
Thus, in  $\bm{P}_{\omega_\gamma}$ and $\bm{M}_{\omega_\gamma}$, only the terms that oscillate at $\omega_{\gamma}$ were retained, see the assumption in Approx. \ref{app:mono}. The neglected oscillatory terms, once substituted into the wave equation, would produce sidebands at frequencies $2 \omega_\gamma \pm \omega_{\ell}$ etc., which can be excluded in an experiment with a spectral filter or spectrometer.
 
Upon substituting the resulting expressions for $\bm{P}_{\omega_\gamma}$ and $\bm{M}_{\omega_\gamma}$ into Eq. \eqref{eq:hemlhotz}, the wave equation can be reduced to a simpler form by making the slowly varying envelope approximation (SVEA - Approx. \ref{app:svea}). This approximation uses the fact that 
\begin{equation}
    \kappa \equiv \text{max}\left(\frac{|\nabla E_{\ell,\gamma}|}{|\omega_{\gamma}E_{\ell,\gamma}|}, \frac{|\partial_tE_{\ell,\gamma}|}{|\omega_{\gamma}E_{\ell,\gamma}|} \right) \ll 1
\end{equation}
to drop higher order derivatives. On the right-hand side of Eq. \eqref{eq:hemlhotz}, the second time derivative of the polarization simplifies to
\begin{equation}
	\begin{split}\label{eq:Pder}
	&4\pi\partial_t^2 \bm{P}_{\omega_\gamma}(t,\bm{x}) =\\ 
 &- \tfrac{1}{2}\omega_\gamma^2 \{\eta_x [E_{\gamma,x}(t,\bm{x}) e^{-i\omega_\gamma(t-z)} + \text{c.c.}] \hb{x}\\
        &+ \eta_y [E_{\gamma,y}(t,\bm{x}) e^{-i\omega_\gamma(t-z)} + \text{c.c.}]\hb{y}\} \\
    &\times \{|E_\ell(t,\bm{x})|^2 + [\tfrac{1}{2}E_\ell^2(t,\bm{x})e^{-2i\omega_\ell(t+z)} + \text{c.c.}]\}\\
    &+ \mathcal{O}(\nu,\kappa)\,,
	\end{split}
\end{equation} 
and similarly for the magnetization
\begin{equation}\label{eq:Mder}
	\begin{split}
 	&4\pi\nabla \times \partial_t \bm{M}_{\omega_\gamma}(t,\bm{x})=\\ 
 &- \tfrac{1}{2}\omega_\gamma^2 \{\eta_x [E_{\gamma,x}(t,\bm{x}) e^{-i\omega_\gamma(t-z)} + \text{c.c.}] \hb{x}\\
        &+ \eta_y [E_{\gamma,y}(t,\bm{x}) e^{-i\omega_\gamma(t-z)} + \text{c.c.}]\hb{y}\} \\
    &\times \{|E_\ell(t,\bm{x})|^2 + [\tfrac{1}{2}E_\ell^2(t,\bm{x})e^{-2i\omega_\ell(t+z)} + \text{c.c.}]\}\\
    &+ \mathcal{O}(\nu,\kappa)\,,
	\end{split}
\end{equation}
which is equal to the polarization term. The third and final term on the right-hand side of Eq. \eqref{eq:hemlhotz} is proportional to $\nabla(\nabla\cdot \bm{P})$ and does not have an $\omega_\gamma^2$ contribution. 

Expressions (\ref{eq:ansatzgamma1},\ref{eq:Pder},\ref{eq:Mder}) are substituted into Eq. (\ref{eq:hemlhotz}), which is broken into two equivalent equations for the non-conjugate and conjugate components. On the left-hand side of Eq. \eqref{eq:hemlhotz}, the terms arising from $\partial_t^2$ and $\partial_z^2$ which are proportional to $\omega_\gamma^2$ cancel, and the remaining second derivative terms are dropped in accordance with the SVEA. What remains is an equation for the amplitude components $E_{\gamma,j}(t,\bm{x})$ where $j \in \{x, y\}$. After multiplying by $2e^{i\omega_\gamma(t-z)}$, one finds
\begin{equation}\label{eq:vector}
	\begin{split}
	[2&i\omega_\gamma (\partial_z + \partial_t) + \nabla_\perp^2]E_{\gamma,j}(t,\bm{x})\\
 &= -2\omega_\gamma^2 \eta_j|E_\ell(t,\bm{x})|^2 E_{\gamma,j}(t,\bm{x})\\
 &-\omega_\gamma^2 \eta_j E_{\gamma,j}(t,\bm{x})[E_\ell^2(t,\bm{x}) e^{-2i\omega_\ell(t+z)} + \text{c.c.}]\\
 &+ \mathcal{O}(\wt{\nu},\kappa)\,.
	\end{split}
\end{equation}
This time $\wt{\nu} \equiv \text{max}(1/\omega_\gamma\hat{w}_\gamma,1/\omega_\ell^2\hat{w}_\ell^2)$ since the longitudinal components proportional to $1/\omega_\ell\hat{w}_\ell$ were dropped. The oscillatory terms in the last expression vanish upon averaging over a laser cycle. Performing a change of variables to the moving frame coordinates $\xi = t - z$ and $\tilde{z} = z$ with the associated derivatives $\partial_t = \partial_\xi$ and $\partial_z = - \partial_\xi + \partial_{\tilde{z}}$ yields the differential equation
\begin{equation}\label{eq:diffeq}
	\begin{split}
	(2i\omega_\gamma &\partial_z + \nabla_\perp^2)E_{\gamma,j}(\xi,\bm{x})\\
	&= - 2 \omega_\gamma^2 \eta_j |E_\ell(\xi,\bm{x})|^2 E_{\gamma,j}(\xi,\bm{x}) + \mathcal{O}(\wt{\nu},\kappa)\,
	\end{split}
\end{equation}
for the components of the complex envelope of the x-ray probe pulse $E_{\gamma,j}(\xi,\bm{x})$, where $\tilde{z}$ has been renamed as $z$. 
\section{FF and conventional laser pulse profiles}
The squared magnitude of the laser field can be expressed as  
\begin{equation}
	|E_\ell(\xi,\bm{x})|^2 = \tilde{E}_\ell^2(\xi,z) g^2_\ell(\xi+2z) e^{-2r^2/w_\ell^2(\xi,z)}\,,
\end{equation}
where $g_\ell(t + z) = g_\ell(\xi+2z)$ is the temporal profile of the pulse and $r\equiv \sqrt{x^2+y^2}$ is the radial distance from the propagation axis. The amplitude and spot size of the FF pulse are given by
\begin{align}
    \label{eq:FFdef1}\tilde{E}_F(\xi,z) &= \tilde{E}_F(\xi) = \frac{\hat{E}_F}{\sqrt{1+(\xi/Z_F)^2}}\,,\\
    \label{eq:FFdef2}w_F(\xi,z) &= w_F(\xi) = \hat{w}_F \sqrt{1+(\xi/Z_F)^2}\,,
\end{align}
where $Z_F \equiv \omega_\ell \hat{w}^2_F$ is the Rayleigh range and $\hat{w}_F$ is the spot size at focus \cite{Palastro_2018, Di_Piazza_2021}. Note the factor of two difference from the standard formula \cite{simpson2022spatiotemporal,Ramsey_2023}. Because the focus moves at the speed of light in the positive $z$ direction, $\tilde{E}_F$ and $w_F$  depend only on $\xi = t - z$. The conventional Gaussian pulse has a stationary focus at $z=0$, such that
\begin{align}
	\label{eq:conventional1}\tilde{E}_G(\xi,z) &= \tilde{E}_G(z) = \frac{\hat{E}_G}{\sqrt{1 + (z/Z_G)^2}}\,,\\
	\label{eq:conventional2}w_G(\xi,z) &= w_G(z) = \hat{w}_G\sqrt{1+(z/Z_G)^2}\,, 
\end{align}
where $Z_G \equiv \omega_\ell \hat{w}^2_G/2$ is the Rayleigh range and $\hat{w}_G$ is the spot size at focus. In this case, both functions depend only on $z$.
\section{Evolution of the x-ray probe pulse} 
Each time slice of the x-ray pulse travels at the speed of light. As a result, the electric field of each time slice can be parameterized by its value of $\xi$ and described by the ansatz
\begin{equation}
	\begin{split} \label{eq:ansatz}
	E_{\gamma,j}&(\xi,\bm{x}) = \tilde{E}_{\gamma,j}(\xi,z)\\
	&\times \exp\left(i \theta_j(\xi,z) - \frac{r^2}{w^2_j(\xi,z)} + i \frac{\omega_\gamma r^2}{2R_j(\xi,z)}\right)\,,
	\end{split}
\end{equation}
where $\tilde{E}_{\gamma,j}(\xi,z)$, $\theta_j(\xi,z)$, $w_j(\xi,z)$, and $R_j(\xi,z)$ are all real functions of $z$. This ansatz has been chosen so that the unknown functions take familiar functional forms in the absence of vacuum-polarization effects, i.e., those of Gaussian optics. Plugging Eq. \eqref{eq:ansatz} into Eq. (\ref{eq:diffeq}), multiplying through by $E_{\gamma,j}^*(\xi,\bm{x})$, and separating the real and imaginary components provides the equations 
\begin{equation}\label{eq:im}
	w^4_j\left(\frac{\tilde{E}'_{\gamma,j}}{\tilde{E}_{\gamma,j}} + \frac{1}{R_j}\right) + 2w_j\left(w_j' - \frac{w_j}{R_j}\right)r^2 = \mathcal{O}(\wt{\nu},\kappa) \,,
\end{equation}
\begin{equation}\label{eq:re}
	\begin{split}
	[\omega_\gamma^2 &w_j^4(R_j'-1)+4R_j^2]r^2 - 2R_j^2w_j^2(2+\omega_\gamma w_j^2\theta_j')\\
	&= - 2\omega_\gamma^2 \eta_j \tilde{E}_\ell^2 g_\ell^2 R_j^2 w_j^4 e^{-2r^2/w_\ell^2}+ \mathcal{O}(\wt{\nu},\kappa)\,,
	\end{split}
\end{equation}
where the prime denotes a partial derivative with respect to $z$, and the dependence of all quantities on $\xi$ and $z$ has been omitted for brevity. Integrating Eqs. \eqref{eq:im} and \eqref{eq:re} over $2 \pi e^{-2r^2/w^2_j} r dr$  yields 
\begin{equation}
	\label{eq:imi1}\frac{\tilde{E}_{\gamma,j}'}{\tilde{E}_{\gamma,j}} = - \frac{w_j'}{w_j} + \mathcal{O}(\wt{\nu},\kappa)\,,\\
\end{equation}
and
\begin{equation}
	\begin{split}
	\label{eq:rei1}\omega_\gamma^2 w_j^4&(R_j'-1)-4R_j^2-4R_j^2\omega_\gamma w_j^2 \theta_j'\\
	 &= - 4 \omega_\gamma^2 \eta_j \tilde{E}_\ell^2 g_\ell^2 R_j^2 w_j^2 \frac{w_\ell^2}{w_j^2 + w_\ell^2}+ \mathcal{O}(\wt{\nu},\kappa)\,.
	\end{split}
\end{equation}
Similarly, integrating Eqs. \eqref{eq:im} and \eqref{eq:re} over $2\pi e^{-2r^2/w^2_j}r^3 dr$ yields
\begin{equation}
	\label{eq:imi2}\frac{w_j'}{w_j} = \frac{1}{R_j} + \mathcal{O}(\wt{\nu},\kappa)
\end{equation}
and
\begin{equation}\label{eq:rei2}
	\begin{split}
	\omega_\gamma^2&w_j^4(R_j'-1)-2R_j^2\omega_\gamma w_j^2 \theta_j'\\
	 &= - 2 \omega_\gamma^2\eta_j \tilde{E}_\ell^2 g_\ell^2 R_j^2 w_j^2 \frac{w_\ell^4}{(w_j^2 + w_\ell^2)^2} + \mathcal{O}(\wt{\nu},\kappa)\,.
	\end{split}
\end{equation} 
Combining Eqs. (\ref{eq:rei1}), (\ref{eq:imi2}), and (\ref{eq:rei2}) provides differential equations for the spot sizes $w_j(\xi,z)$ and phase shifts $\theta_j(\xi,z)$ of the x-ray probe pulse
\begin{align}
w''_j - \frac{4}{\omega_\gamma^2 w^3_j}\left[1 - \omega_\gamma^2 \eta_j \tilde{E}_\ell^2 g_\ell^2 \frac{w_\ell^2 w^4_j}{(w^2_j + w_\ell^2)^2}\right] &=\mathcal{O}(\wt{\nu},\kappa)\label{eq:exactw}\,,\\
\theta'_j + \frac{2}{\omega_\gamma w^2_j} - \omega_\gamma \eta_j \tilde{E}_\ell^2 g_\ell^2 \frac{w_\ell^2 (2w^2_j + w_\ell^2)}{(w^2_j + w_\ell^2)^2} &=\mathcal{O}(\wt{\nu},\kappa)\label{eq:exacttheta}\,.
\end{align}
In addition, Eqs. \eqref{eq:imi1} imply conservation of  power,  $w_j^2(\xi,z) \tilde{E}_{\gamma,j}^2(\xi,z) \approx $ constant, and the radii of curvature $R_j(\xi,z)$ can be found from Eq. \eqref{eq:imi2} once $w_j(\xi,z)$ is known. 

\section{Perturbative solution}\label{sec:perturbative}
Equations \eqref{eq:exactw} and \eqref{eq:exacttheta} can be solved perturbatively by expanding in orders of the small dimensionless parameter $\eta_j\hat{E}_\ell^2$ (Approx. \ref{app:weak}). The spot sizes and phases are written as 
\begin{align}
	w_j(\xi,z) &= w^{(0)}(z) + \delta w_j(\xi,z)\,,\\
	\theta_j(\xi,z) &= \theta^{(0)}(\xi,z) + \delta\theta_j(\xi,z)\,.
\end{align}
The zeroth-order solutions are equal to the spot size and phase of a Gaussian beam (Approx. \ref{app:gaussian}) in the absence of vacuum polarization effects:
\begin{align}
	w^{(0)}(z) &= \hat{w}_\gamma \sqrt{1+z^2/Z_\gamma^2}\,,\label{eq:w0}\\
	\theta^{(0)}(\xi,z) &= \theta_0(\xi) - \arctan(z/Z_\gamma)\,,
\end{align}
where $\theta_0(\xi)$ is an arbitrary real function and $Z_\gamma \equiv \omega_\gamma \hat{w}_\gamma^2/2$. Equations \eqref{eq:imi1} and \eqref{eq:w0} imply that the zeroth-order amplitudes $\tilde{E}^{(0)}_j(\xi,z) \propto 1/w^{(0)}(z)$ depend only on $z$. 

The corrections $\delta w_j(\xi,z)$ and $\delta \theta_j(\xi,z)$ arise from  vacuum-polarization effects. To first order in $\eta_j\hat{E}_\ell^2$ (Approx. \ref{app:weak}), the corrections satisfy the differential equations
\begin{equation}
    \begin{split}
	\delta w_j'' + \frac{12\delta w_j}{\omega_\gamma^2(w^{(0)})^4} + 4\eta_j \wt{E}_\ell^2 g_\ell^2 w_\ell^2 \frac{w^{(0)}}{[(w^{(0)})^2 + w_\ell^2]^2}\\
    = \mathcal{O}(\wt{\nu},\kappa,\varepsilon^2) \,,
    \end{split}\label{eq:numw}
\end{equation}
\begin{equation}
    \begin{split}
	\delta \theta_j' - \frac{4\delta w_j}{\omega_\gamma (w^{(0)})^3} - \omega_\gamma \eta_j \wt{E}_\ell^2 g_\ell^2 w_\ell^2 \frac{2(w^{(0)})^2 + w_\ell^2}{[(w^{(0)})^2 + w_\ell^2]^2}\\
    = \mathcal{O}(\wt{\nu},\kappa,\varepsilon^2)\,,
    \end{split}\label{eq:numtheta}
\end{equation}
where $\varepsilon \equiv \hat{E}_\ell^2 /E_\text{cr}^2$. This system of equations can be used to find numerical solutions for $\delta w_j(\xi,z)$ and $\delta \theta_j(\xi,z)$. In order to derive analytical estimates for these quantities, the equations will be solved in the region where $z^2/Z_\gamma^2 \ll 1$ (Approx. \ref{app:short}). Further, the temporal profiles of the laser pulses will be approximated as rectangles, i.e., $g_{\ell} = 1$ for the duration of the interaction and $g_{\ell} = 0$ otherwise (Approx. \ref{app:rect2}).

\subsection{FF pulse}
For the FF pulse, $\tilde{E}_F(\xi)$ and $w_F(\xi)$ are given by Eqs. \eqref{eq:FFdef1} and \eqref{eq:FFdef2}. When the length of the x-ray pulse $L$ is much smaller than the interaction length $D_F$, the initial conditions $\delta w_j(-D_F/2) = \delta w_j'(-D_F/2) = 0$, $\delta \theta_j(-D_F/2) = 0$ can be imposed. The leading order solution for $z \in (-D_F/2,D_F/2)$ is then
\begin{equation}
    \begin{split}
	\delta w_j^F(\xi,z) &= - 2 \eta_j \tilde{E}_F^2(\xi) \frac{\hat{w}_\gamma w_F^2(\xi)}{[\hat{w}_\gamma^2 + w_F^2(\xi)]^2}\\
    &\times \left(z+\frac{D_F}{2}\right)^2
    + \mathcal{O}(\wt{\nu},\kappa,\varepsilon^2,d^3)\,,
    \end{split}
 \end{equation}
 \begin{equation}\label{eq:FFphase}
    \begin{split}
	\delta \theta_j^F(\xi,z) &= \omega_\gamma \eta_j \tilde{E}_F^2(\xi) \frac{w_F^2(\xi)[2\hat{w}_\gamma^2 + w_F^2(\xi)]}{[\hat{w}_\gamma^2 + w_F^2(\xi)]^2}\\
    &\times \left(z+\frac{D_F}{2}\right)
    + \mathcal{O}(\wt{\nu},\kappa,\varepsilon^2,d^2)\,,
    \end{split}
\end{equation}
where higher order terms in $d \equiv D_\ell/Z_\gamma$ were neglected (Approx. \ref{app:short}). Thus in obtaining this solution, the terms proportional to $\delta w_j$ in Eqs. (\ref{eq:numw}) and (\ref{eq:numtheta}) were omitted. By substituting the solution back into these equations, one can verify that this is a valid approximation consistent with Approx. \ref{app:short}. 

The vacuum-polarization nonlinearity depends on polarization, i.e., it is birefringent. This results in a different $\delta\theta$ for each polarization component of the x-ray pulse. The difference in these phase shifts, $\Delta \theta_F = \delta \theta_y - \delta \theta_x$, provides a measurable signature of the birefringence. Specifically,  
\begin{equation}\label{eq:analytical}
	\begin{split}
    \Delta \theta_F (\xi,z) &= \frac{2\alpha}{15} \left(\frac{\tilde{E}_F(\xi)}{E_\text{cr}}\right)^2 \frac{w_F^2(\xi)[2\hat{w}_\gamma^2 + w_F^2(\xi)]}{[\hat{w}_\gamma^2 + w_F^2(\xi)]^2} \\
    &\times \frac{z+D_F/2}{\lambda_\gamma} + \mathcal{O}(\wt{\nu},\kappa,\varepsilon^2,d^2)\,.
    \end{split}
\end{equation}
The value of $\xi$ determines the synchronization of a particular time slice of the x-ray pulse with respect to the peak intensity of the flying-focus pulse. For the temporal slice of the x-ray pulse that is colocated with the intensity peak of the FF, $\xi = 0$, and this expression reduces to
\begin{equation}\label{eq:FFsynced}
    \begin{split}
    \Delta \theta_F (0,z) &= \frac{2\alpha}{15} \left(\frac{\hat{E}_F}{E_\text{cr}}\right)^2  \frac{z+D_F/2}{\lambda_\gamma}\Sigma_F\\
    &+ \mathcal{O}(\wt{\nu},\kappa,\varepsilon^2,d^2)\,,
    \end{split}
\end{equation}
where the transverse overlap factor $\Sigma_F$ is given by
\begin{equation}
\Sigma_F \equiv \frac{1 + 2\sigma_F^2}{(1 + \sigma_F^2)^2}\,
\end{equation}
and $\sigma_F\equiv \hat{w}_\gamma/\hat{w}_F$. The phase difference increases as $\sigma_F \rightarrow 0$. However, if this limit is achieved by focusing the x-ray pulse too tightly, the approximation $D_F/Z_\gamma \ll 1$ would not be valid. Nevertheless, for $\sigma_F \ll 1$ and $z=D_F/2$ the phase difference reduces to
\begin{align}
    \begin{split}
	\Delta \theta_F \left(0,\frac{D_F}{2}\right) = \frac{2\alpha}{15} \left(\frac{\hat{E}_F}{E_\text{cr}}\right)^2 \frac{D_F}{\lambda_\gamma}\\
    + \mathcal{O}(\wt{\nu},\kappa,\varepsilon^2,d^2)\,,
    \end{split}
\end{align}
which matches the result for the interaction of a probe with a strong plane wave field over a length $D_F$ (see, e.g., Ref. \cite{Dittrich:2000zu}).

Experimentally, it is infeasible to temporally resolve the polarization of the x-ray pulse after its interaction with the FF pulse. This motivates averaging Eq. (\ref{eq:analytical}) over all $\xi$ within the x-ray pulse:
\begin{equation}\label{eq:averaged}
	\begin{split}
        \Delta \theta_F&(z) \equiv \frac{1}{L}\int_{-L/2+d}^{L/2 + d} \Delta \theta_F(\xi,z) d\xi\\ &= \frac{2\alpha}{15} \left(\frac{\hat{E}_F}{E_\text{cr}}\right)^2 \frac{z + D_F/2}{\lambda_\gamma}\Sigma_F\Lambda_F(L,d)\\
        &+ \mathcal{O}(\wt{\nu},\kappa,\varepsilon^2,d^2)\,.
	\end{split}
\end{equation}
This is an exact result for the approximate expression in Eq. (\ref{eq:analytical}). The longitudinal form-factor $\Lambda_F(L,d)$ depends on the initial synchronization and geometry of the pulses and is given by
\begin{equation}\label{eq:LambdaF}
		\begin{split}
        \Lambda_F&(L,d) \equiv \frac{1}{2}\frac{\rho_F/L}{ 1+2\sigma_F^2}\\
        &\times \left[\frac{\sigma^2_F u}{1 + u^2} + (2+3\sigma_F^2)\arctan u \right]_{\frac{-L/2+d}{\rho_F}}^{\frac{L/2+d}{\rho_F}}\,.
        \end{split}
\end{equation}
In Eq. \eqref{eq:LambdaF}, $L$ is the length of the x-ray probe pulse, $d$ is the longitudinal displacement of its center from the peak intensity of the FF pulse, and $\rho_F \equiv Z_F\sqrt{1+\sigma_F^2}$. With $d = 0$ and in the limit as $L \rightarrow 0$, $\Lambda_F(L,d) \rightarrow 1$. 

%

\subsection{Conventional Gaussian pulse}
For the conventional laser pulse, $\tilde{E}_G(z)$ and $w_G(z)$ are given by Eqs. \eqref{eq:conventional1} and \eqref{eq:conventional2}. In this case, the length of the x-ray pulse $L$ is comparable to the interaction length $D_G$ and must be accounted for explicitly. A best case scenario of perfect synchronization is assumed so that the centers of both pulses meet at the focus of the conventional pulse $z=0$. This means that leading and trailing edges of both pulses meet symmetrically around the focus at $z=-D_G/2 + L/4$ and $z = D_G/2 - L/4$, respectively. This ensures that each temporal slice of the x-ray pulse interacts with the laser pulse in the vicinity of its focus over the entire interaction length. The temporal slices of the x-ray pulse are again parameterized by $\xi \in (-L/2,L/2)$ but this time with no offset. The leading order solution of Eq. \eqref{eq:numtheta} after the entire interaction is then
\begin{equation}
	\begin{split}
	\delta \theta_j^G(\xi) = &\frac{1}{2}\eta_j \hat{E}_G^2 \frac{\omega_\gamma Z_G}{(1 + \sigma_G^2)^{3/2}}\\
	&\times \left[\frac{\sigma_G^2 u}{1 + u^2} + (2 + 3 \sigma_G^2)\arctan u\right]_{u_i(\xi)}^{u_f(\xi)}\\
    &+ \mathcal{O}(\wt{\nu},\kappa,\varepsilon^2,d^2)\,,
	\end{split}
\end{equation}
where $\rho_G \equiv Z_G\sqrt{1+\sigma^2_G}$, $\sigma_G \equiv \hat{w}_\gamma / \hat{w}_G$, and 
\begin{equation}
    u_f(\xi) \equiv \frac{D_G/2 + \xi/2}{\rho_G}\,,\quad
    u_i(\xi) \equiv \frac{-D_G/2 + \xi/2}{\rho_G}.
\end{equation}
As in the FF case, the terms proportional to $D_G/Z_\gamma$ (Approx. \ref{app:short}), i.e., $\delta w_j$ in Eqs. (\ref{eq:numw}) and (\ref{eq:numtheta}) were neglected. The phase difference $\Delta \theta_G = \delta \theta_y - \delta \theta_x$ is given by
\begin{equation}\label{eq:analyticalF}
	\begin{split}
    \Delta \theta_G(\xi) &=  \frac{\alpha}{15}\left(\frac{\hat{E}_G}{E_\text{cr}}\right)^2\frac{Z_G}{\lambda_\gamma(1 + \sigma_G^2)^{3/2}}\\
    & \times \left[\frac{\sigma_G^2 u}{1 + u^2} + (2 + 3 \sigma_G^2)\arctan u\right]_{u_i(\xi)}^{u_f(\xi)}\\
    &+ \mathcal{O}(\wt{\nu},\kappa,\varepsilon^2,d^2)\,.
    \end{split}
\end{equation}
After averaging over all $\xi$ within the x-ray pulse, one finds
\begin{equation}\label{eq:gaussian}
	\begin{split}
    \Delta \theta_G &= \frac{1}{L}\int_{-L/2}^{L/2} \Delta \theta_G(\xi)d\xi \\ &= \frac{2\alpha}{15}\left(\frac{\hat{E}_G}{E_\text{cr}}\right)^2
    \frac{2\rho_G}{\lambda_\gamma} \Sigma_G \tilde{\Lambda}_G(L,D_G)\\
    &+ \mathcal{O}(\wt{\nu},\kappa,\varepsilon^2,d^2)\,,
    \end{split}
\end{equation}
which is an exact result for the approximate expression in Eq. (\ref{eq:analyticalF}). Here, the transverse and longitudinal form factors are
\begin{align}
    \Sigma_F &\equiv \frac{1+2\sigma_F^2}{(1+\sigma_F^2)^2}\,,\\
    \label{eq:LambdaGtilde}\tilde{\Lambda}_G(L,D_G) &\equiv \frac{\rho_G}{L} \frac{1}{1 + 2\sigma_G^2} [W(u)]^{u_f(L/2)}_{u_f(-L/2)}
\end{align}
and the auxiliary function 
\begin{equation}
    W(u) \equiv (2 + 3 \sigma_G^2)u\arctan u - (1 + \sigma_G^2) \ln(1+u^2)\,.
\end{equation}
In the limit of an ultrashort x-ray pulse ($L \rightarrow 0$) and for small interaction lengths ($D_G\ll Z_G$),  $\tilde{\Lambda}_G \approx D_G/2\rho_G$.  
\section{Phase difference as a function of laser pulse energy}\label{sec:app_const_energy}
In this appendix, the factors $\hbar$, $\varepsilon_0$, and $c$ are made explicit for clarity. The cycle-averaged power of a laser pulse with a Gaussian transverse profile is given by 
\begin{equation}\label{eq:Pave}
	P_{\ell} = \frac{\pi}{4}\varepsilon_0 c \hat{E}_{\ell}^2\hat{w}_{\ell}^2+ \mathcal{O}(1/\omega_\ell^2\hat{w}_\ell^2)\,.
\end{equation}
The energy of the pulse is $\mathcal{E}_{\ell} = \tau_{\ell}P_{\ell}$, where $\tau_{\ell}$ is the duration.

The duration of the FF pulse is determined by the focal range $D_F$ and velocity of the focus $v_F$: $\tau_{F} = |c^{-1}-v_F^{-1}|D_F$ \cite{simpson2022spatiotemporal,Formanek_2022,Ramsey_2023}. For $v_F = -c$, 
\begin{equation}
	\mathcal{E}_F = \frac{2D_F}{c}P_F = \frac{\pi}{2} \varepsilon_0 \hat{E}_F^2 D_F \hat{w}_F^2 + \mathcal{O}(1/\omega_F^2\hat{w}_F^2)\,.
\end{equation}
Substituting $\hat{E}_F^2$ into Eq. \eqref{eq:averaged}, setting $z=D_F/2$, and averaging over the x-ray frequency spectrum yields 
\begin{equation}\label{eq:FFresult}
    \begin{split}
	\Delta \theta_F  &= \frac{8\alpha^2}{15\pi}\frac{\mathcal{E}_F}{e^2 E_\text{cr}^2} \frac{\hbar \langle \omega_\gamma \rangle}{\hat{w}_F^2}\Sigma_F \Lambda_F(L,d)\\
    &+ \mathcal{O}(\wt{\nu},\kappa,\varepsilon^2,d^2)\,.
    \end{split}
\end{equation}
The phase difference is independent of the focal range $D_F$ and linearly proportional to the laser pulse energy. Given an energy and a focal range, the average power can be calculated as
\begin{equation}
	P_F = \frac{c\mathcal{E}_F}{2D_F},
\end{equation}
which does not depend on the spot size $\hat{w}_F$. 

The duration of the conventional pulse is set to ensure that the x-ray pulse and conventional pulse overlap over the entire interaction length, i.e., $\tau_{G} = 2D_G/c$. The energy of the pulse is then 
\begin{equation}
	\mathcal{E}_G = \frac{2D_G}{c} P_G = \frac{\pi}{2}\varepsilon_0 \hat{E}_G^2 D_G\hat{w}_G^2+\mathcal{O}(1/\omega_G^2\hat{w}_G^2)\,.
\end{equation}
Substituting $\hat{E}_G^2$ into Eq. \eqref{eq:gaussian} and averaging over the x-ray frequency spectrum yields
\begin{equation}\label{eq:Gresult}
    \begin{split}
	\Delta \theta_G &= \frac{8\alpha^2}{15\pi}\frac{\mathcal{E}_G}{e^2 E_\text{cr}^2} \frac{\hbar \langle \omega_\gamma \rangle}{\hat{w}_G^2}\Sigma_G\Lambda_G(L,D_G)\\
    &+ \mathcal{O}(\wt{\nu},\kappa,\varepsilon^2,d^2)\,,
    \end{split}
\end{equation}
where the form factor $\Lambda_G$ contains an additional coefficient that depends on the interaction length $D_G$
\begin{equation}\label{eq:LambdaG}
    \Lambda_G(L,D_G) \equiv \frac{2\rho_G}{D_G}\tilde{\Lambda}_G(L,D_G)\,.
\end{equation}
Other than the form factors $\Lambda_\ell$, the phase difference for a given pulse energy is identical in the FF [Eq. (\ref{eq:FFresult})] and conventional cases [Eq. (\ref{eq:Gresult})]. Given an energy and an interaction length, the average power of the conventional pulse can be calculated as
\begin{equation}
	P_G = \frac{c\mathcal{E}_G}{2D_G}\,.
\end{equation}    
Typically, the interaction length should be comparable to the Rayleigh range. For the purposes of Fig. \ref{fig:power_ratio}, $D_G = Z_G$.

\section{Numerical implementation}\label{sec:app_numerical}
\begin{figure}
    \centering
    \includegraphics[width=\linewidth]{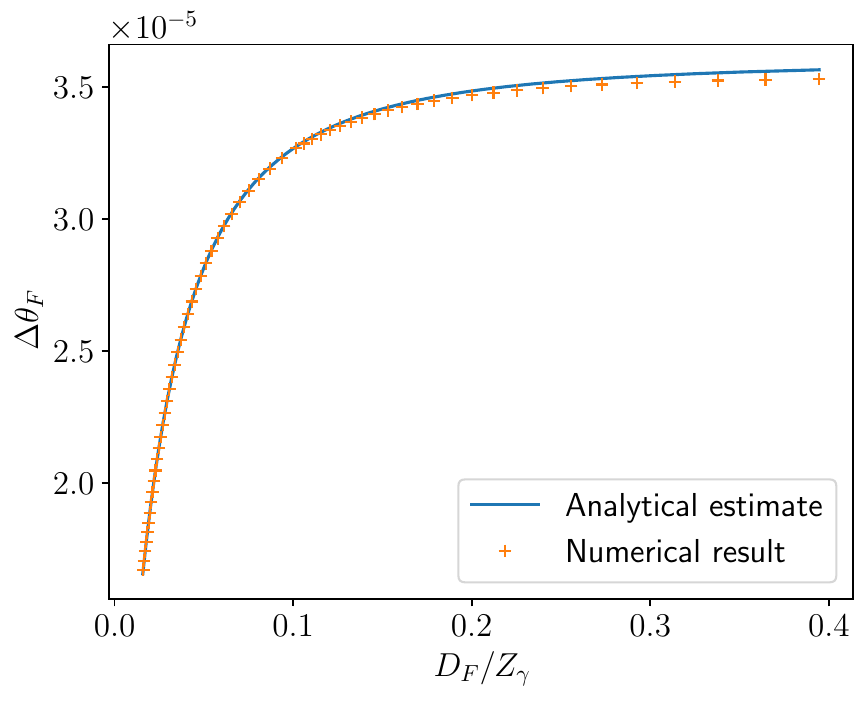}
    \caption{\label{fig:num_vs_analyt}Phase difference for a short, perfectly synchronized 10 keV x-ray pulse due to vacuum birefringence in a FF pulse with $\lambda_F = 1\ \mu$m, $\hat{w}_F = 3\ \mu$m, $\mathcal{E}_F = 1000$ J, and $P_F = 15$ TW. The interaction length is $D_F = 1$ cm. The phase difference is plotted as a function of $D_F/Z_\gamma$ for the numerical solution [solving Eqs. \eqref{eq:numw} and \eqref{eq:numtheta} for both polarizations] and the analytical approximation in Eq. \eqref{eq:FFresult}. The range of $D_F/Z_\gamma$ values corresponds to spot sizes $\hat{w}_\gamma \in (1,5)\ \mu$m.}
\end{figure}
In order to numerically solve Eqs. \eqref{eq:numw} and \eqref{eq:numtheta} in the FF case, the envelope function $g_F$ needs to be specified. Here a smooth polynomial function is employed, which is more realistic version of the rectangular pulse profile used for the analytical estimates. Specifically,
\begin{equation}
    g_F(z) = \left\{\begin{matrix}
    10\tilde{z}_+^3 - 15\tilde{z}_+^4 + 6\tilde{z}_+^5, & \tilde{z}_+ 
    \in (0,1)\,,\\
    1, & \tilde{z}_+ \in [1,\tilde{D}_F] \,,\\
    -10\tilde{z}_-^3-15\tilde{z}_-^4-6\tilde{z}_-^5,& \tilde{z}_- 
    \in (-1,0),\\
    0, & \text{otherwise}\,,
\end{matrix}\right.
\end{equation}
where $\tilde{z}_\pm \equiv (z \pm D_F/2 \pm L_r/2)/L_r$, $\tilde{D}_F \equiv D_F/L_r$, and $L_r$ is the ramp length. The length of the ramps are  set to be 0.5\% of the interaction length. This ensures that the ramps are long compared to the wavelength of the laser, but short compared to the overall pulse duration. 

The system of Eqs. \eqref{eq:numw} and \eqref{eq:numtheta} was solved using the 4th order Runge-Kutta integration scheme \cite{press2007numerical} with a step $\Delta z = 0.1\ \mu$m. The phase difference $\Delta \theta_F$ was calculated by subtracting the solutions for each polarization $\Delta \theta_F = \delta \theta_y^F - \delta \theta_x^F$.

As was discussed in Sec. \ref{sec:perturbative}, the approximation allowing for an analytical solution begins to break down when the Rayleigh range of the x-ray pulse $Z_{\gamma}$ becomes comparable to the interaction length $D_F$. To determine the accuracy of the analytical results, the case of a perfectly synchronized ($d = 0$), short ($L \ll Z_F$) 10 keV x-ray probe pulse was simulated using Eqs. \eqref{eq:numw} and \eqref{eq:numtheta}. In the simulations, the phase accumulated over an interaction length of 1 cm in a vacuum polarized by a FF pulse with a spot size $\hat{w}_F = 3\ \mu$m and an energy $\mathcal{E}_F = 1$ kJ. Figure \ref{fig:num_vs_analyt} shows the phase difference predicted by the analytical result [solid line, Eq. \eqref{eq:FFresult}] and the numerical integration (discrete points) as a function of the small parameter $D_F / Z_\gamma$. The results are in excellent agreement for the range of x-ray spot sizes considered, $\hat{w}_{\gamma} =$ 1 - 5 $\mu$m. The small discrepancy for tightly focused x-ray pulses disappears if higher x-ray photon energies are considered (thus increasing $Z_F$). Moreover, focusing the x-ray probe pulse much tighter than the laser pulse does not appreciably modify the phase difference (see discussion in the main text). 

%

\end{document}